\documentclass[a4paper,fleqn,usenatbib]{mnras}

\usepackage[T1]{fontenc}
\usepackage{ae,aecompl}

\usepackage{graphicx}	
\usepackage{amsmath}	
\usepackage{amssymb}	

\newcommand{\Msun}{$M_\odot$}

\title[PISN models versus SLSNe]
{Can pair-instability supernova models match \\
the observations of superluminous supernovae?}

\author[A. Kozyreva et al.]{
Alexandra Kozyreva,$^{1, 2, 3}$\thanks{E-mail: a.kozyreva@keele.ac.uk}
S. Blinnikov$^{4, 5, 6}$
\\
$^{1}$Astrophysics group, Keele University, Keele, Staffordshire, ST5 5BG, UK \\
$^{2}$Argelander-Institut f\"ur Astronomie, Universit\"at Bonn, Bonn, 53121, Germany \\
$^{3}$Sternberg Astronomical Institute, Lomonosov Moscow State University, Moscow, 119992, Russia \\
$^{4}$ITEP (Kurchatov Institute), Moscow, 117218, Russia \\
$^{5}$VNIIA, Moscow, 127473, Russia\\
$^{6}$Kavli IPMU (WPI), University of Tokyo, Japan
}

\date{Accepted XXX. Received YYY; in original form ZZZ}

\pubyear{2015}

\begin{document}
\label{firstpage}
\pagerange{\pageref{firstpage}--\pageref{lastpage}}
\maketitle

\begin{abstract}
An increasing number of so-called superluminous supernovae (SLSNe) are discovered.
It is believed that at least some of them with slowly fading light curves originate in
stellar explosions induced by the pair instability mechanism.  
Recent stellar evolution models naturally predict pair instability supernovae (PISNe)
from very massive stars at wide range of metallicities (up to $Z =
0.006$, Yusof~et~al.~2013). 
In the scope of this study we analyse whether PISN models can match the observational properties of SLSNe
with various light curve shapes.  Specifically,
we explore the influence of different degrees of macroscopic chemical mixing in PISN
explosive products on the resulting observational properties.
We artificially apply mixing to the 250~\Msun{}~PISN evolutionary model from Kozyreva\,et\,al.\,(2014) and
explore its supernova evolution with the one-dimensional radiation hydrodynamics code STELLA.
The greatest success in matching SLSN observations is achieved in the case
of an extreme macroscopic mixing, where all radioactive material is ejected 
into the hydrogen-helium outer layer.  
Such an extreme macroscopic redistribution of chemicals produces events with 
faster light curves with high photospheric
temperatures and high photospheric velocities.  These properties fit a wider 
range of SLSNe than non-mixed PISN model.
Our mixed models match the light curves, colour temperature and photospheric
velocity evolution of two well-observed SLSNe PTF12dam and LSQ12dlf.  
However, these models' extreme chemical redistribution may be hard
to realise in massive PISNe.  Therefore, alternative models such as
the magnetar mechanism or wind-interaction may still to be favourable to interpret rapidly rising SLSNe.
\end{abstract}




\section[Introduction]{Introduction}
\label{sect:intro}

Thanks to the launch of an increasing number of supernova surveys,
the number of superluminous supernovae (hereafter, SLSNe; see \citet[][]{2012Sci...337..927G} for a review) grows 
\citep[see
e.g.,][]{2009Natur.462..624G,2011Natur.474..487Q,2012Natur.491..228C,2014MNRAS.441..289B,2014MNRAS.437..656M,2014MNRAS.444.2096N,2015MNRAS.452.3869N}.  
The extreme property for all of them is high peak luminosity, and many show blue featureless spectra at the discovery.  
Some of them have slow decline, while others have more rapid decline.

There are three possible mechanisms currently suggested for SLSNe.  These are 
(1) central engine, such as magnetar or accreting black hole 
\citep[see e.g.,][]{2006Natur.442.1018M,2010ApJ...719L.204W,2010ApJ...717..245K,2013ApJ...772...30D}, 
(2) interaction-powered mechanism
\citep{2011ApJ...729L...6C,2012ApJ...757..178G,2013MNRAS.430.1402M,2015AstL...41...95B}, and 
(3) nickel-powered mechanism \citep{2009Natur.462..624G,2010ApJ...717L..83M,2012Sci...337..927G}.  
In the current paper we focus on supernovae powered by radioactive nickel-cobalt decay, 
which occurs in pair instability supernovae (hereafter, PISNe).

Stars with initial masses above 150~\Msun{} definitely exist in the visible Universe
\citep{2008MNRAS.389L..38S,2010MNRAS.408..731C,2014ApJ...780..117S}.  
Their evolution is fairly clear and successfully reproduced by stellar evolution simulations 
\citep[][and others]{1964ApJS....9..201F,1967SvA....10..604B,1967ApJ...148..803R,1967PhRvL..18..379B,1968Ap&SS...2...96F,1986A&A...167..274E,2002ApJ...567..532H,2002ApJ...565..385U,2007A&A...475L..19L,2012ApJ...760..154C,2013MNRAS.433.1114Y,2014A&A...566A.146K}.  
Many of these studies predict that these stars form 
massive oxygen core (above 60~\Msun{}) which eventually explodes due to pair creation instability mechanism.
In theory, explosions of these very massive stars should be detected among other supernova explosions.  
According to the relative number of progenitors, one pair-instability explosion of a very massive star 
is expected for every one thousand core-collapse explosions of massive stars \citep{2007A&A...475L..19L,2008A&A...489..359Y}.  
Nevertheless, the explosion of PISN even for a low-mass progenitor 
($\sim$~150~\Msun{}) appears sufficiently bright to be detected at
large distances \citep{2011ApJ...734..102K,2014A&A...565A..70K}.  
Hence we presume that observationally the number of PISNe is higher than proposed ratio 1:1000.


The main property of high mass PISN explosions is a very 
broad light curve \citep{2011ApJ...734..102K,2013MNRAS.428.3227D,2014A&A...565A..70K}, because during the explosion
the entire progenitor is expelled into the surrounding space.  The massive ejecta, up to 200~\Msun{}, cause 
a very long diffusion time.  Therefore, the rise to the peak lasts up to 200~days.  
The shallow decline follows the nickel radioactive decay.  
In contrast to this, many SLSNe have faster light curve evolution.

If radioactive material is ejected into the upper envelope, then the rise time to the maximum shortens.  
As a consequence, the ejecta is more compact at peak epoch, and hence colour
temperature\footnote{Colour temperature is a temperature of a black body
spectrum which is close to the spectral density distribution, i.e. continuum spectrum.} 
is higher.  Besides that, nickel additionally heats the layers in which it is distributed 
\citep[see e.g.,][]{1990ApJ...360..242S,2004ApJ...617.1233Y,2013ApJ...767..143B}.  
In any case, ejecting freshly produced nickel from the innermost region is challenging
and difficult to model in one-dimensional stellar evolution codes.

The present successful multi-dimensional simulations of PISN explosions 
do not reveal extended mixing \citep{2011ApJ...728..129J,2014ApJ...792...44C}.  However,
the observed spectra reveal a presence of metals (including iron) in the spectra of SLSNe 
\citep{2009Natur.462..624G,2013Natur.502..346N,2014MNRAS.444.2096N} soon after maximum 
(30\,--\,50~days later).  This motivates a toy experiment assuming different amounts of mixing for our
original high-mass PISN model, as explained below.  Additionally,
we analyse the dependence of the final light curve shape on  different degree of the envelope stripping.

It is well-known that the famous SN~1987A light curve requires an extensive mixing of radioactive nickel
into the helium-hydrogen atmosphere \citep{1990ApJ...360..242S,1993A&A...270..249U}.
Nevertheless, computer simulations hardly reproduce mixing of 
nickel in SN~1987A \citep[see e.g.,][and references therein]{2009ApJ...693.1780J}.  
To explain the high nickel velocities observed in SN~1987A many studies suggest clumping and ejection of the innermost 
hot matter into the overlying shells \citep[see e.g.,]{1989ARA&A..27..629A,1990ApJ...360..257H,1994ApJ...425..264B}.  
\citet{1989Natur.341..489C} suggests that low-density bubbles could arise as a result of the high-entropy interplay.  
Simulation by \citet{1988PASJ...40..691N} and \citet{1991ApJ...367..619F} show the naturally emerging 
fragmentation occurring on the early stages of a supernova explosion. 

Microscopic diffusive chemical mixing occurs in core-collapse supernovae (hereafter, CCSNe) basically due to 
the Richtmyer--Meshkov instability, which is similar to the Rayleigh--Taylor instability 
left behind the passage of the shock wave \citep{2009ApJ...693.1780J}.  
A number of studies reveal also that high velocity macroscopic blobs of metal-rich material could appear
during neutrino-induced CCSNe
\citep[see e.g.,][and references therein]{2006A&A...453..661K,2010ApJ...714.1371H}.  
The reasons for macro-mixing in CCSNe are: (1) proto-neutron star convection, (2) neutrino-driven convection, 
and (3) magneto-rotational instability \citep[see for discussion][and references therein]{2014ApJ...793...45N}.  
The preferred direction for buoyant-bubble growth lies towards the equatorial plane.

Even though the physics of PISN differs from that of CCSNe.  We propose that there is some possibility that
entropy perturbations occur during PISN explosion.  
The interplay between high-entropy bubbles arising from the explosive site and
cooler overlying layers is principal feasible in PISNe, 
because the pair creation and subsequent explosive burning occur at sufficiently high entropy.  
Explosive oxygen and subsequently silicon burning provides further increase in entropy.  
The alpha-disintegration occurring at the very centre
lowers entropy causing heterogeneous features in isentropic structure.
Hence, macro-mixing or strong anisotropy might take place in PISN ejecta
instead of micro-mixing.  However, we note that bulk motion of massive macroscopic blobs (10--20~\Msun{}) is hard to realise in the
explosive time-scale.

We describe our toy models in Section~\ref{sect:method}, discuss the resulting light curves and photospheric evolution in
Section~\ref{sect:results}.  
In Section~\ref{sect:discussion}, we explain applicability of our models to SLSNe.  
We conclude our study in Section~\ref{sect:conclusions}.

\section[Input models and light curves modelling]{Input models and light curves modelling}
\label{sect:method}

Our original stellar evolution model is taken from \citep{2014A&A...566A.146K}.  
The particular interest in the scope of this study lies in the high-mass PISN model (with initial mass 250~\Msun{}).  
A high-mass PISN produces a very large amount of radioactive nickel \citep{2002ApJ...567..532H} 
which powers the supernova light curve.  The resulting supernova appears superluminous 
reaching maximum luminosity up to several $10^{\,44}$~erg\,s$^{\,-1}$ (absolute magnitude up to --\,22~mag).  
Our 250~\Msun{} PISN generates 19.3~\Msun{} of nickel and radiates $10^{\,44}$~erg\,s$^{\,-1}$ at peak luminosity.

\begin{figure}
\centering
\includegraphics[width=0.5\textwidth]{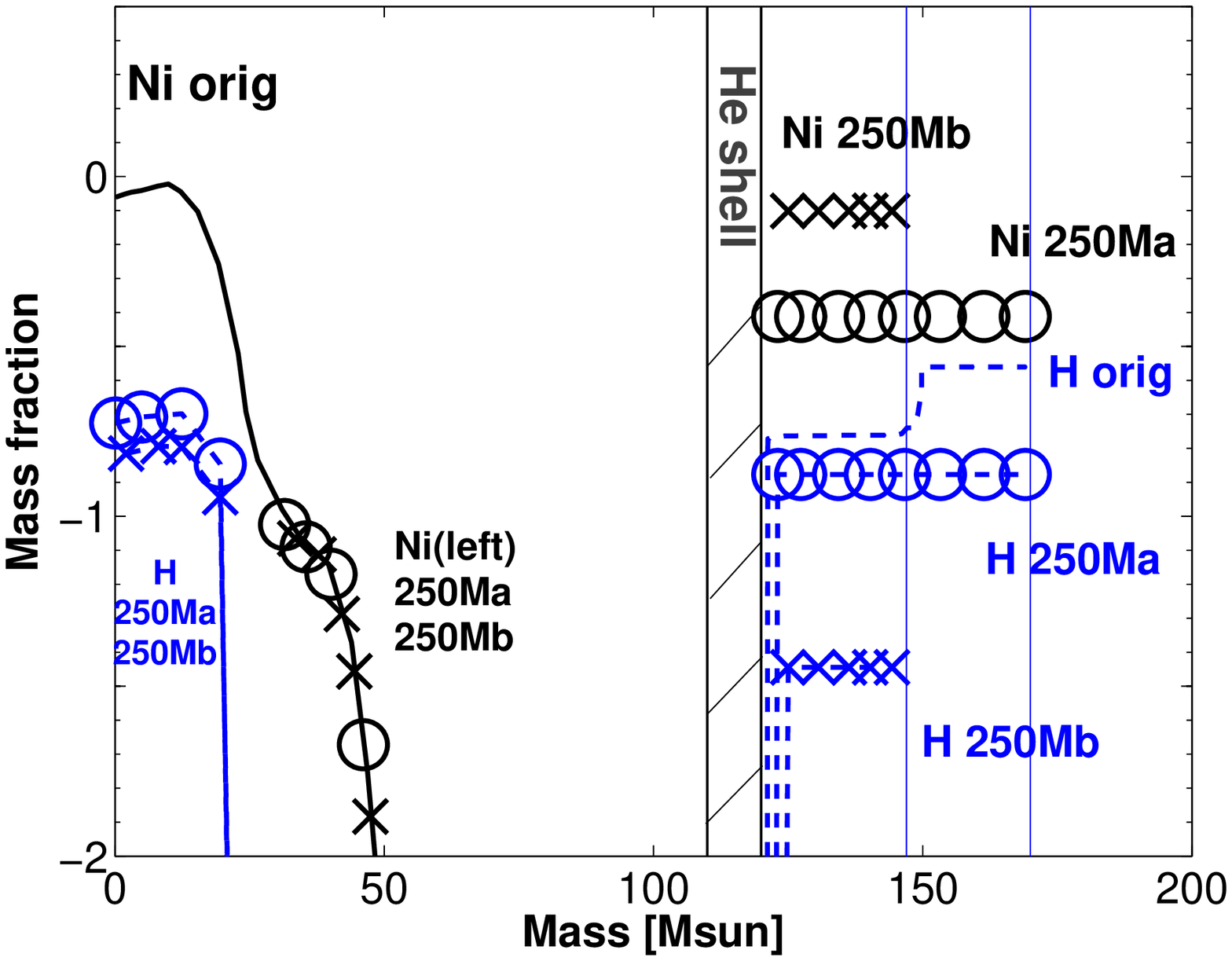}\\
\includegraphics[width=0.5\textwidth]{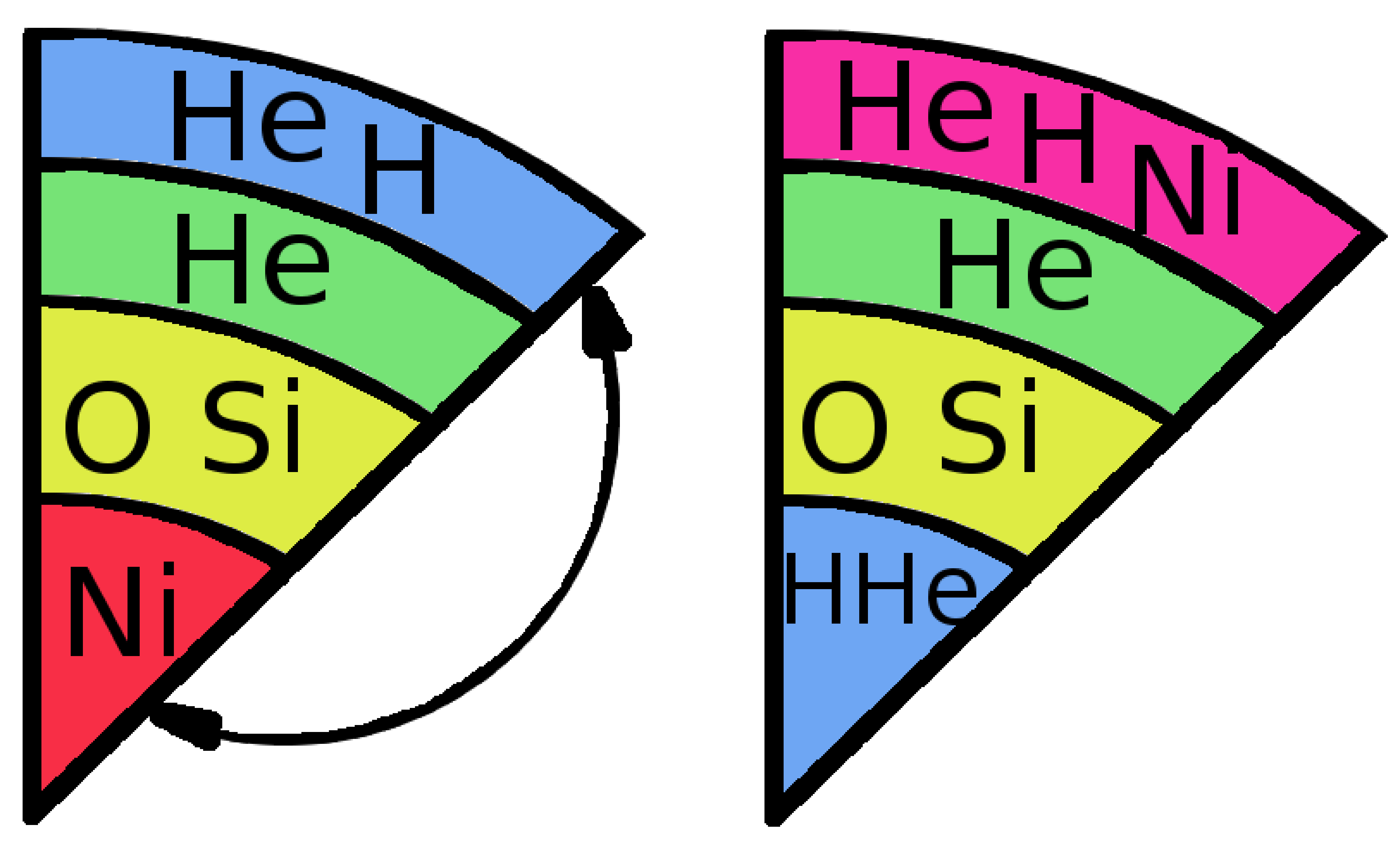}
\caption[Distribution of nickel, hydrogen and helium in the input PISN models]
{{\textit{Top}}: Distribution of nickel (black) and hydrogen (blue) in 
the original model 250M (solid and dashed lines), and two mixed models 250Ma
(circles) and 250Mb (time signs).  Hatched region between 110~\Msun{} and
120~\Msun{} indicates the location of helium shell.  Thin line at
147~\Msun{} shows the outer boundary of model~250Mb, and that is at
170~\Msun{} shows the outer boundary of model~250M (original) and
model~250Mb.
\\
{\textit{Bottom}}: Illustration explaining redistribution of chemical elements
in the mixed toy models.  In the mixed models we replaced nickel from the
centre into the outer H/He atmosphere.  }
\label{figure:chemie}
\end{figure}

Note, that our PISN model evolves in a self-consistent way from the zero-age main sequence, 
follows the pair-creation phase, undergoes explosive oxygen and silicon burning, and eventually explodes.  
The calculations were carried out with the stellar evolution code BEC with the extended nuclear network
\citep{2007A&A...475L..19L,2006A&A...460..199Y,2014PhDTKozyreva}.  
We address the reader to our earlier paper \citep{2014A&A...566A.146K}
for all details about the evolution and explosion of the PISNe.

Metallicity of our model is 0.001 which is lower than those of hosts of SLSNe 
2007bi and PTF12dam \citep{2010A&A...512A..70Y,2015MNRAS.452.1567C}.  
Slightly higher metallicity leads to higher mass loss and absence of
hydrogen in the outer atmosphere of the final PISN progenitor \citep{2013MNRAS.433.1114Y}.  
The uncertainty of mass-loss rate allows us to predict the evolutionary model of
a very massive star retaining hydrogen atmosphere.  
Nevertheless, we emphasise that our study is qualitative and demonstrate
what kind of observational properties a high-mass hydrogen-rich PISN
may have and how the properties change if the model is modified.  
On top of that, we suspect that metallicity is not the same for the entire galaxy and might be
different for the supernova site compared to the averaged value.  

\subsection[Effect of stripping]{Effect of stripping}
\label{subsect:cut}

As an initial attempt we ran a number of supernova models for our 250M PISN, in which we subsequently
strip the hydrogen-helium envelope.  Hence we have a sequence of the following stripped models: 
152~\Msun{}, 139~\Msun{}, 132~\Msun{}, and 127~\Msun{}.  Note, that our original model 250M 
contains 169~\Msun{} at the moment of pair-instability explosion.  
The sequence 152, 139, 132~\Msun{} roughly corresponds to the mass of truncated outermost shell --- 20~\Msun{},
30~\Msun{}, 40~\Msun{}.  
In the model 127~\Msun{}, the edge lies just above the helium
shell, so this model closely corresponds to the 130~\Msun{} helium model \citep{2011ApJ...734..102K}.

We do not change the chemical composition in these models.  We append a tiny stratified 
atmosphere to satisfy the outer boundary condition (vanishing pressure).  The composition of the atmosphere is the same as 
in the upper layer, at which we truncate the model.

\subsection[Effect of mixing]{Effect of mixing}
\label{subsect:mix}

We inspect different kinds of hypothetical ``mixing''.  
We emphasise here that we do not assume microscopic convective mixing operating through the diffusive processes.  
The pair-instability explosion develops on a dynamical timescale, which is considerably shorter than the convective time.  

Based on the multi-dimensional numerical simulations, convective time in the carbon-burning
convective shell during core oxygen burning in a massive star model is about 100\,s \citep{2013ApJ...769....1V}.  
The convective oxygen core roughly has the same convective timescale.  
Quantitatively, convective time could be estimated as
$t_{\rm{conv}} = 2 \left({G M\over{\rho r^2}} \Delta\nabla\rho \right)^{-1/2}$
\citep{2002RvMP...74.1015W}.  This gives an approximate value about 
10\,seconds in the core.  The convective timescale gradually increases up to
$10^{\,5}$~s above the oxygen core.  However, mentioned timescale is
related to regular convection connected to hydrostatic carbon/oxygen
burning.  It is not excluded that convection might be accelerated on the on-going
core-collapse and subsequent explosion.  Dynamical time corresponds to a free-fall time
$t_{\rm{ff}}\sim (G\rho)^{-1/2}$.  Hence dynamical time is less than second in the core of our
PISN model and getting comparable to convective time in the envelope. 

To conclude, we exclude the relevance of microscopic mixing taking place
during pair-instability explosion.  Everywhere in the present study we refer to macroscopic displacement of stellar matter
as ``mixing'' without meaning microscopic convective mixing.  Below we detail our selected attempts.

\subsubsection[Intermediate mixing]{Intermediate mixing}
\label{subsubsect:mix1}

In view of the results of recent studies \citep{2011ApJ...728..129J,2014ApJ...792...44C}, we calculated a couple of 
explosions with intermediate mixing.  Under ``intermediate'' we mean the mixing happens 
in the intermediate oxygen layers.  
\citet{2011ApJ...728..129J} and \citet{2014ApJ...792...44C} claim that
possible mixing can occur at the oxygen-helium interface of extended red supergiants.
We ran explosions for models in which we mixed material contained in the mass layers between 
(1) 60~\Msun{} and 130~\Msun{}, and (2) 80~\Msun{} and 150~\Msun{} along Lagrangian mass coordinate 
\footnote{As \citet{2014ApJ...792...44C} say, the model 225~\Msun{} manifests 
``complete destruction of the helium and oxygen layer and partly silicon shell''.}.  
The case (1) implies uniform mixing of the oxygen and helium shells.  In the case (2), we propose mixing between 
half of the oxygen layer ($\sim 20$~\Msun{}), the complete helium shell
($\sim 10$~\Msun{}) and part of the hydrogen-helium atmosphere ($\sim 30$~\Msun{}).

However, we show in Section~\ref{subsect:halfres} that none of the models with intermediate mixing produce light curves
very different from the original unmixed PISN model.  This happens because
the hydrogen and nickel distribution is not modified, although hydrogen and nickel
remain the most crucial in formation of the light curve \citep{1993A&A...270..249U}.

\subsubsection[Uniform mixing]{Uniform mixing}
\label{subsubsect:mix2}

Next, we present a few toy models that were computed assuming 
(1) an overall uniform mixing of the entire star, 
(2) an uniform distribution of nickel through the inner regions of the star inside 120~\Msun{},
i.e. up to the upper edge of the oxygen layer, and
(3) an uniform distribution of nickel in 30~\Msun{} at the upper boundary of oxygen shell.  
In the later model we additionally depleted the hydrogen-helium atmosphere by 10~\Msun{}.  
As we show in Section~\ref{subsect:halfres}, even though these 
models produce light curves which significantly differ from the
original model supernova evolution, they poorly match SLSN properties.

\subsubsection[Extreme nickel mixing]{Extreme nickel mixing}
\label{subsubsect:mix}

Finally, we calculate an extraordinary toy experiment motivated thus.  The main properties of
some SLSNe are fast rise time, high colour temperature and
high photospheric velocity.  These properties are hardly reproduced by massive PISN models.  It is well-known that faster rise to
the peak luminosity can be achieved by putting radioactive material into the upper layers
\citep{1990ApJ...360..242S,2014ApJ...784...85P}.  
Another consequence of this relocation is increase in photospheric temperature.

For extreme mixing we relocated almost all of the radioactive nickel from the innermost 20~\Msun{} region
to the hydrogen-helium envelope.  To fulfil total mass conservation we displaced
outermost 20~\Msun{} of hydrogen-helium into the centre.  This is done only
because of simplification for realisation of our toy model.  Otherwise,
outer 20~\Msun{} of hydrogen and helium might be decayed amongst intermediate layers. 
However, the light curve will be strongly affected by hydrogen recombination in this case
\citep{1990ApJ...360..242S,1993A&A...270..249U}.  
Such extreme chemical displacement sounds difficult
to realise, but provides light curve evolution suitable for explaining properties of SLSNe.  

Our modified chemical structures are shown in Figure~\ref{figure:chemie}
(see the bottom panel of Figure~\ref{figure:chemie} for illustration).  In the \textbf{model~250Ma}, 
almost all radioactive nickel was replaced into the outer hydrogen-helium layer.  
Vice versa, the same amount of hydrogen and helium (with the original proportion of mass fraction 20:80) 
was replaced into the centre.  In the \textbf{model~250Mb}, 
firstly we artificially removed 20~\Msun{} of the outer
hydrogen-helium atmosphere, and then displace nickel and hydrogen/helium (similarly to the model~250Ma).

\vspace{1cm}

The calculations of the explosion evolution is carried out with 
the one-dimensional multigroup radiation Lagrangian implicit hydrodynamics code {\sc STELLA}
(see \citealt{2006A&A...453..229B}, \citealt{2014A&A...565A..70K} for details and references therein).  

It is clear that a one-dimensional code treats the proposed ``mixing'' as uniform. 
An uniform mixing might be a result of micro-mixing processes.  
However, diffusive microscopic mixing of 20~\Msun{} or 100~\Msun{} of stellar matter operates on the timescale
much longer than dynamical timescale on which the explosion occurs.  At the same time matter can be ejected 
in the form of macroscopic clumps or blobs without a requirement of mixing with the ambient stellar matter. 
Nevertheless, this is difficult to model with a one-dimensional code.  
We discuss this point in the following sections.

\section{Results}
\label{sect:results}

\begin{figure}
\centering
\includegraphics[width=0.5\textwidth]{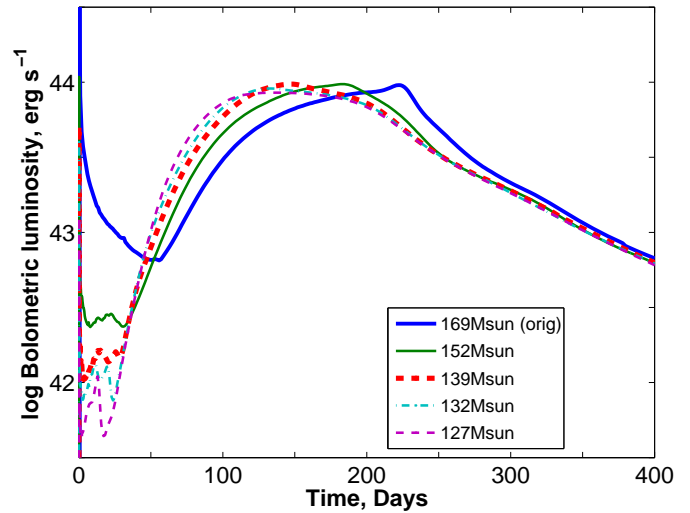}
\caption[]{Bolometric theoretical light curves for the original PISN model 250M (blue
thick solid), and for models with subsequently reduced hydrogen-helium envelope: 152~\Msun{} (green
thin solid), 139~\Msun{} (red thick dashed), 132~\Msun{} (cyan 
dashed-dotted), and 127~\Msun{} (magenta thin dashed).  See discussion in the text.}
\label{figure:Lbol_cut}
\end{figure}

\subsection[Halfway results]{Halfway results}   
\label{subsect:halfres}

In Figure~\ref{figure:Lbol_cut}, we show resulting light curves for the \textbf{truncated models}.  Shrinking of the hydrogen-helium
envelope does not lead to significant changes in the light curve width\footnote{Diffusion time mainly depends on the ejecta mass 
($t_{\mathrm{diff}}\sim E^{\,-1/4} M^{\,3/4} \varkappa^{\,1/2}$) \citep{1977ApJS...33..515F,2009ApJ...703.2205K}.  
Hence, decrease in mass by 40~\Msun{} reduces the overall diffusion time by less than 25\%.}.  
The shape of the photospheric phase slightly changes:
rise to maximum becomes sharper, and the nickel-powered photospheric phase becomes more symmetric (dome-like).  

Noticeable difference is related to the phase between shock breakout and the rise to maximum.  The model~250M has the
longest shock-cooling phase, because the model contains the largest envelope in the sense of radius.  Once the
tenuous part of the envelope is depleted, the outermost layer relaxes very quickly after the shock breakout event.  
At the same time stripping of the envelope causes the plateau-like phase between shock breakout and re-brightening to the
nickel-powered maximum.  This phase is governed by recombination in
the helium shell, so that the light curve is a result of recession of recombination
front combined with a overall expansion of the ejecta.  In model~127~\Msun{} the light curve even has a
prominent local maximum during this intermediate phase.

We conclude that the main benefit of stripped models is the decrease in the rise time.  Light curves of these models rise
slightly faster to maximum than initial model, keeping the overall duration long enough.  
Generally speaking, evolutionary models of very massive stars with inclusion of higher mass-loss rates and/or rotation might
result in PISN progenitors lacking the massive hydrogen-helium atmosphere
and part of helium shell \citep{2013MNRAS.433.1114Y,2015ApJ...799...18C}.  
Therefore, we think that future evolutionary calculations of rotating very massive stars might be relevant for SLSNe.  
Nevertheless, we emphasise that truncation of the hydrogen-helium envelope in our original model~250M does not lead to desirable changes in observational
signatures of PISNe and does not provide sufficiently successful results in explaining SLSNe.  We expect more significant changes
for mixed models, which we describe in the following paragraphs.  

\begin{figure}
\centering
\includegraphics[width=0.5\textwidth]{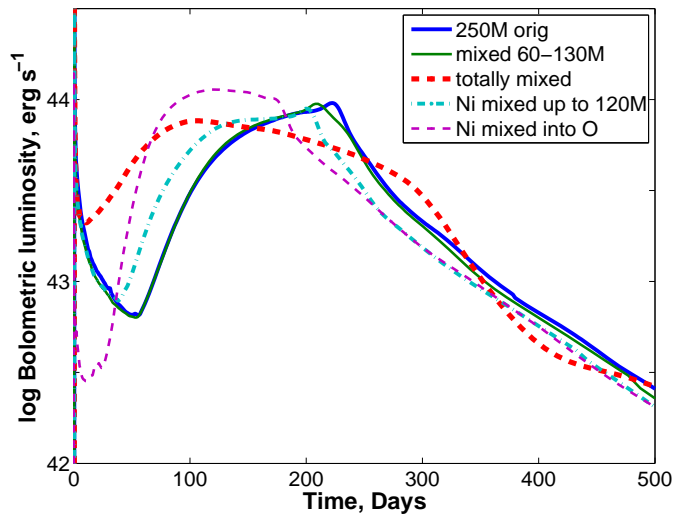}
\caption[]{Bolometric theoretical light curves for the original PISN model 250M (blue
thick solid), and modified mixed models: matter is uniformly mixed between 60~\Msun{} and 130~\Msun{} (green
thin solid), fully uniformly mixed model (red thick dashed), nickel is uniformly mixed up to helium layer (cyan
dashed-dotted), the innermost 30~\Msun{} is replaced with the uppermost 30\Msun{} of oxygen layer (magenta
thin dashed).  
See discussion in the text.}
\label{figure:Lbol_diffMix}
\end{figure}

In Figure~\ref{figure:Lbol_diffMix}, we demonstrate light curves for intermediately and uniformly mixed models.  
The light curve for the \textbf{original PISN model~250M} is labelled ``250M orig'' and appears as a
blue line.  

\textbf{Uniform intermediate mixing} of oxygen and helium shells (between 60~\Msun{} and 130~\Msun{} along 
mass coordinate) does not provide big a difference from the original light curve.  This is mostly because the modified 
model (green line labelled ``mixed 60-130M'') has no changes in hydrogen and nickel distribution compared to the model~250M.  
Both hydrogen and nickel are the principal species that govern dynamics of electron-scattering photosphere.

\textbf{Totally mixed model} is presented by the red line (labelled ``totally mixed'').  
Since nickel is distributed up to the edge of the star,
it starts powering the resulting light curve earlier than model~250M.  Therefore, nickel-powered maximum occurs at 
day~100, while it happens at day~220 in the original model 250M.  Hydrogen being throughout 
the ejecta prevents photosphere from recession and
makes the light curve very broad, even broader than the original one.  Because hydrogen retains in the outer layer, 
the slope of rise to the peak luminosity is similar to model~250M.

\textbf{Smearing nickel throughout the inner regions} allows re-brightening to occur earlier 
(model labelled ``Ni mixed up to 120M'').  However, since overall chemical structure (except nickel) is the same as
model~250M, there is no other difference between the light curves.

\textbf{Displacement of all nickel with 30~\Msun{} at the upper boundary of oxygen shell}
significantly modifies the shape of the original light curve.  
It sharply rises to ``maximum plateau'' (shown in magenta, labelled ``Ni mixed into O'').  
Such short rise time is explained by distribution of nickel in a narrow shell far from the centre. 
The explanation for the plateau-like shape is the following.  
The high energy photons from nickel decay heat the
overlying layer and keep the hydrogen ionised, preventing recession of the electron-scattering photosphere.  
The width of the light curve is slightly less than model~250M, because
10~\Msun{} of the outer atmosphere were depleted.

To conclude the halfway results, we suggest the following.  To force the light curve to rise faster, 
nickel has to appear close to the edge of the ejecta and be distributed locally in the mass coordinate.  
To reduce the light curve width, the hydrogen mass fraction in the ejecta ideally should be reduced.

\subsection[Main results]{Main results}
\label{subsect:mainres}

\begin{figure}
\centering
\includegraphics[width=0.5\textwidth]{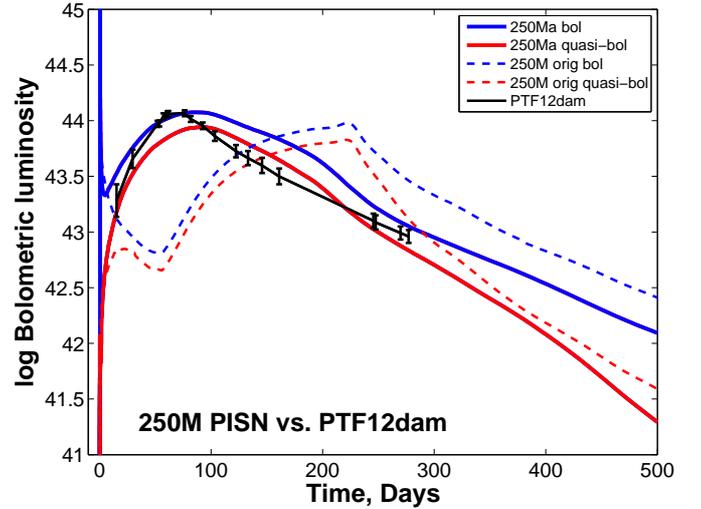}
\caption[]{Bolometric (blue) and quasi-bolometric (red) theoretical light curves for the original PISN 
model~250M (dashed) and mixed PISN model~250Ma and bolometric light curve of SLSN PTF12dam 
(black, \citet{2013Natur.502..346N})}
\label{figure:Lbol1}
\end{figure}

Figures~\ref{figure:Lbol1} and \ref{figure:Lbol2} show bolometric and
quasi-bolometric\footnote{For the quasi-bolometric light curve we integrate the flux
between 3250\AA{} and 8900\AA{}. } light curves for the 
original unmixed PISN model 250M \citep{2014A&A...565A..70K}, and mixed PISN models 250Ma and 250Mb.  
The observed quasi-bolometric light curves of SLSNe PTF12dam and LSQ12dlf are also superposed
\footnote{We chose these two particular supernovae because among other SLSNe they have relatively broad 
light curves, so that PISN models are able to match them.  
In addition, complete observational data for these supernovae were available during the time of the 
present numerical experiment.}
\citep{2013Natur.502..346N,2014MNRAS.444.2096N}.  The observed curves are shifted, so that 
the computed light curves roughly follow the observations during rise, maximum and
post-maximum epochs.  PTF12dam data are shifted on 80~days for luminosity,
colour temperature and photospheric velocity plots,
and LSQ12dlf data are shifted on 60~days.

We see that the original light curve is very broad and indeed encounters 
difficulty to match observed narrower light curves.  Luminosity rises to the peak value during 200~days, while the observed
luminosity increases during 50~days and less for PTF12dam and LSQ12dlf, respectively.  
However, redistribution of radioactive material strongly modifies the 
radiative properties of the PISN explosion, so that the PISN models 250Ma and 250Mb could explain
the observational appearance of SLSNe.  We note, however, that
other SLSNe have even narrower light curves with very short rise 
time and sharp decay after maximum.  
The magnetar nature or circumstellar interaction is probably the best for explaining these events 
\citep{2010ApJ...717..245K,2010ApJ...719L.204W,2012MNRAS.426L..76D,2013ApJ...773...76C,2015MNRAS.452.3869N}.

It is known that many SLSNe have blue spectra soon after their discovery and hence high colour temperatures
\citep{2014MNRAS.441..289B,2013Natur.502..346N,2014MNRAS.444.2096N}.  
All PISN models in \citet{2013MNRAS.428.3227D} have cool photospheres and low temperatures.  
As \citet{2014A&A...565A..70K} show, colour temperature of our original PISN model~250M is 
higher during maximum phase than hydrogen recombination temperature.  This is because maximum luminosity happens when the photosphere
already leaves the hydrogen-rich layer.  Nevertheless, the colour temperature of model~250M hardly explains majority of blue SLSNe.

\begin{figure*}
\centering
\includegraphics[width=0.5\textwidth]{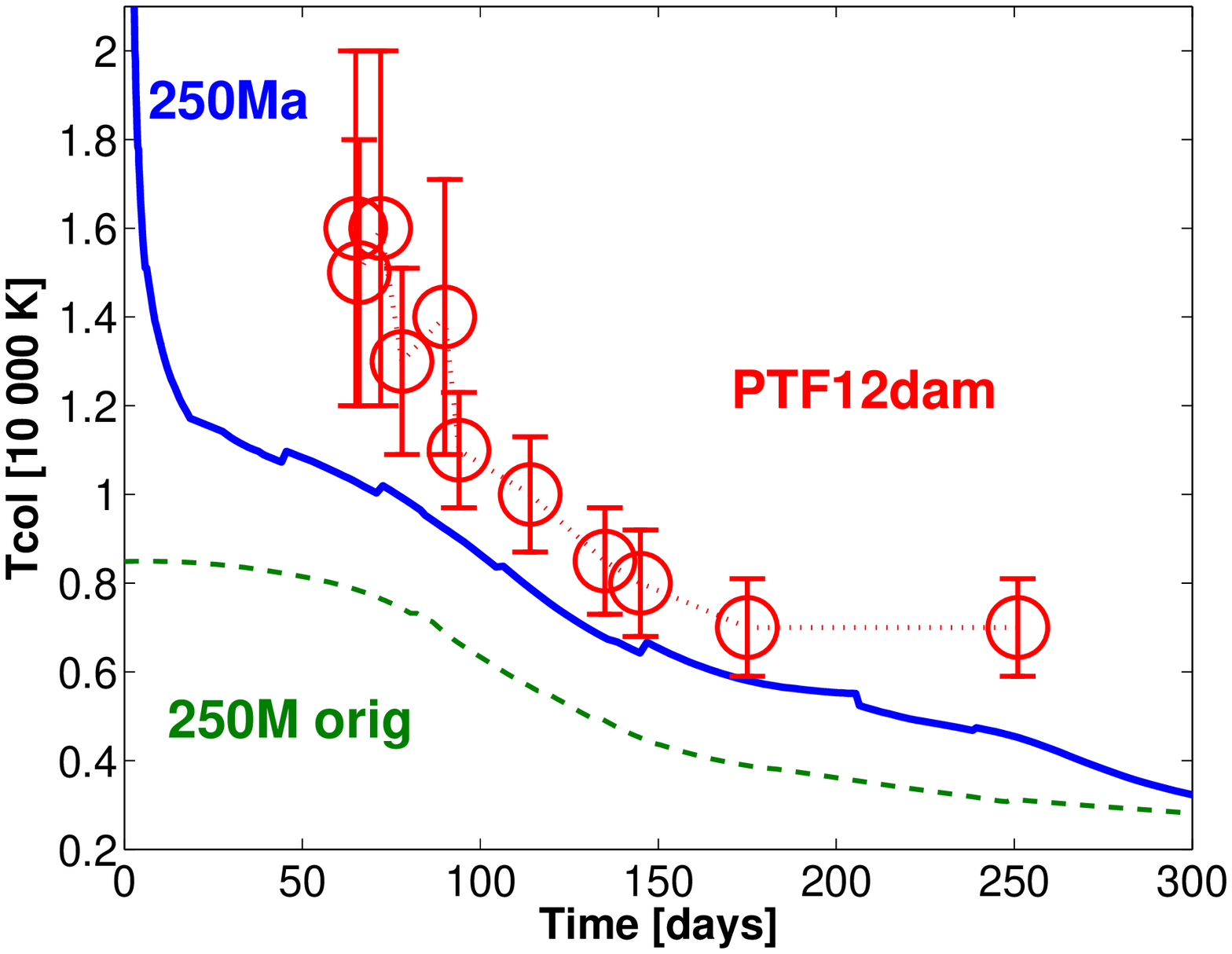}~
\includegraphics[width=0.5\textwidth]{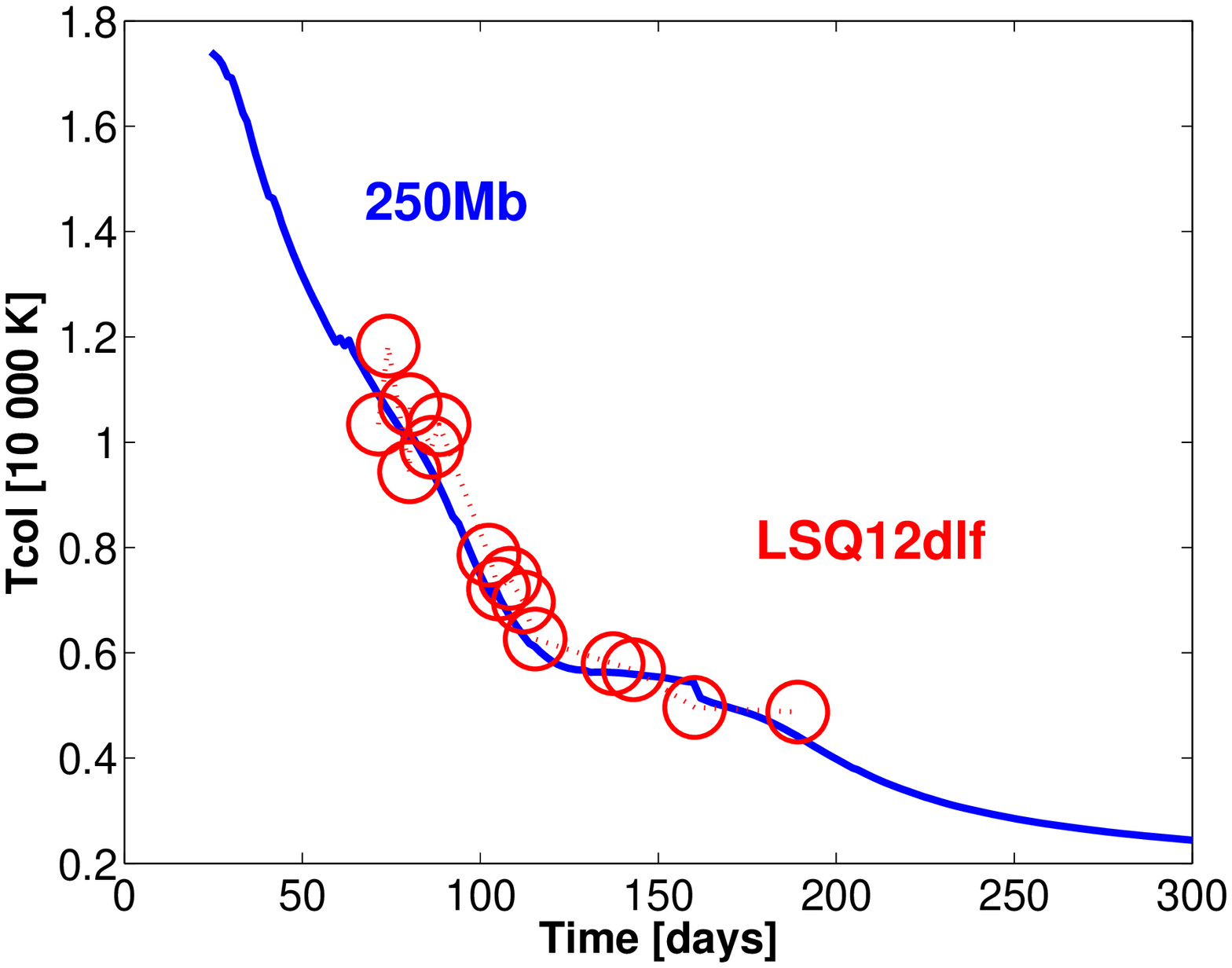}
\caption[]{{\emph{Left}}: 
Colour temperature of the original PISN model~250M (dashed line), the mixed model 250Ma (solid thick line) and 
observed SLSN PTF12dam (red circles with error-bars, shifted on 80~days, \citet{2013Natur.502..346N});\\
{\emph{Right}}:
Colour temperature of the mixed model 250Mb (solid thick line) and observed SLSN
LSQ12dlf (red circles, shifted on 60~days, \citet{2014MNRAS.444.2096N}).  See explanation in the text.}
\label{figure:Tcol}
\end{figure*}

We find out that extended mixing strongly enhances colour
temperature\footnote{We estimate colour temperature based on the least-square method using the
spectral range from 1\AA{} to 50 000\AA{}.} around supernova maximum.  
We show the results and comparison to observations in Figure~\ref{figure:Tcol}.  
The ejecta is more compact and hotter for models~250Ma and 250Mb at the time of supernova maximum, 
because it occurs significantly earlier compared to the unmixed model.  
In Figures~\ref{figure:Tcol} and \ref{figure:Uph} (top), the theoretical
curves of the original model~250M is shown with a
shift --130~days.  PTF12dam temperature and velocity data are shifted on 80~days, similarly to the
light curve comparison.  LSQ12dlf data are equally shifted on 60~days for
luminosity, temperature and velocity comparison in corresponding figures.  
As Figure~\ref{figure:Tcol} shows, the observed temperature of SLSN~PTF12dam 
is still slightly higher than those of model~250Ma.  However, we presume that the observed temperature
estimate might contain a significant uncertainty.

Similarly to photospheric temperature many SLSNe demonstrate high photospheric velocities \citep{2014MNRAS.444.2096N}.  
This property makes them resemble supernovae~Type~Ic.  
On the contrary, \citet{2014A&A...565A..70K} and \citet{2013MNRAS.428.3227D} show that massive PISN ejecta 
provide relatively low velocities, about 5000~km\,s$^{\,-1}$.  
We find that radioactive material appearing in the upper layers effectively changes the 
photospheric velocity evolution (see Figure~\ref{figure:Uph}).  
Ejecting radioactive material into the outer layers provides 
an earlier maximum, therefore, the photosphere is located in the faster layers at earlier time.  
Note, that many SLSNe have featureless spectra at
early epochs, which encounters difficulties in estimating accurate photospheric velocities.

Even though our PISN model is hydrogen-helium rich, the maximum occurs when the 
receding electron-scattering photosphere already left the massive hydrogen-helium 
envelope and moves along a thick oxygen shell.  
Therefore, if the supernova is discovered at maximum, it might look like a SN~Ic, but the presence
of observable H/He signatures can not be ruled out.

\section{Discussion}
\label{sect:discussion}

\begin{figure}
\centering
\includegraphics[width=0.5\textwidth]{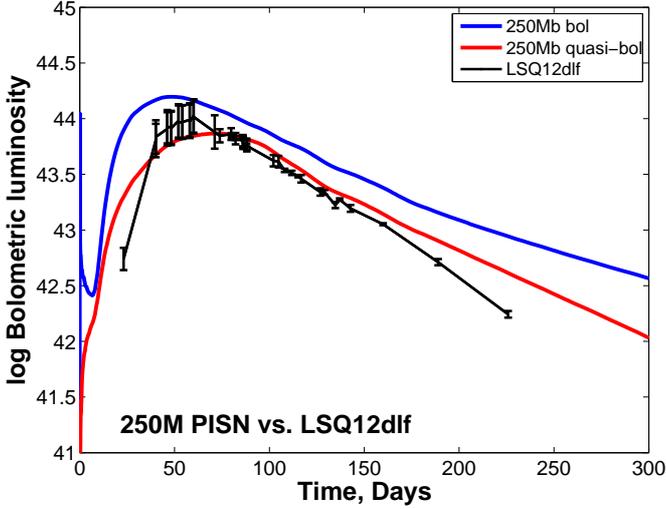}
\caption[]{Bolometric (blue) and quasi-bolometric (red) theoretical light curves for the 
mixed PISN model 250Mb and bolometric light curve of SLSN LSQ12dlf (black, \citet{2014MNRAS.444.2096N})}
\label{figure:Lbol2}
\end{figure}

\begin{figure*}
\centering
\includegraphics[width=0.5\textwidth]{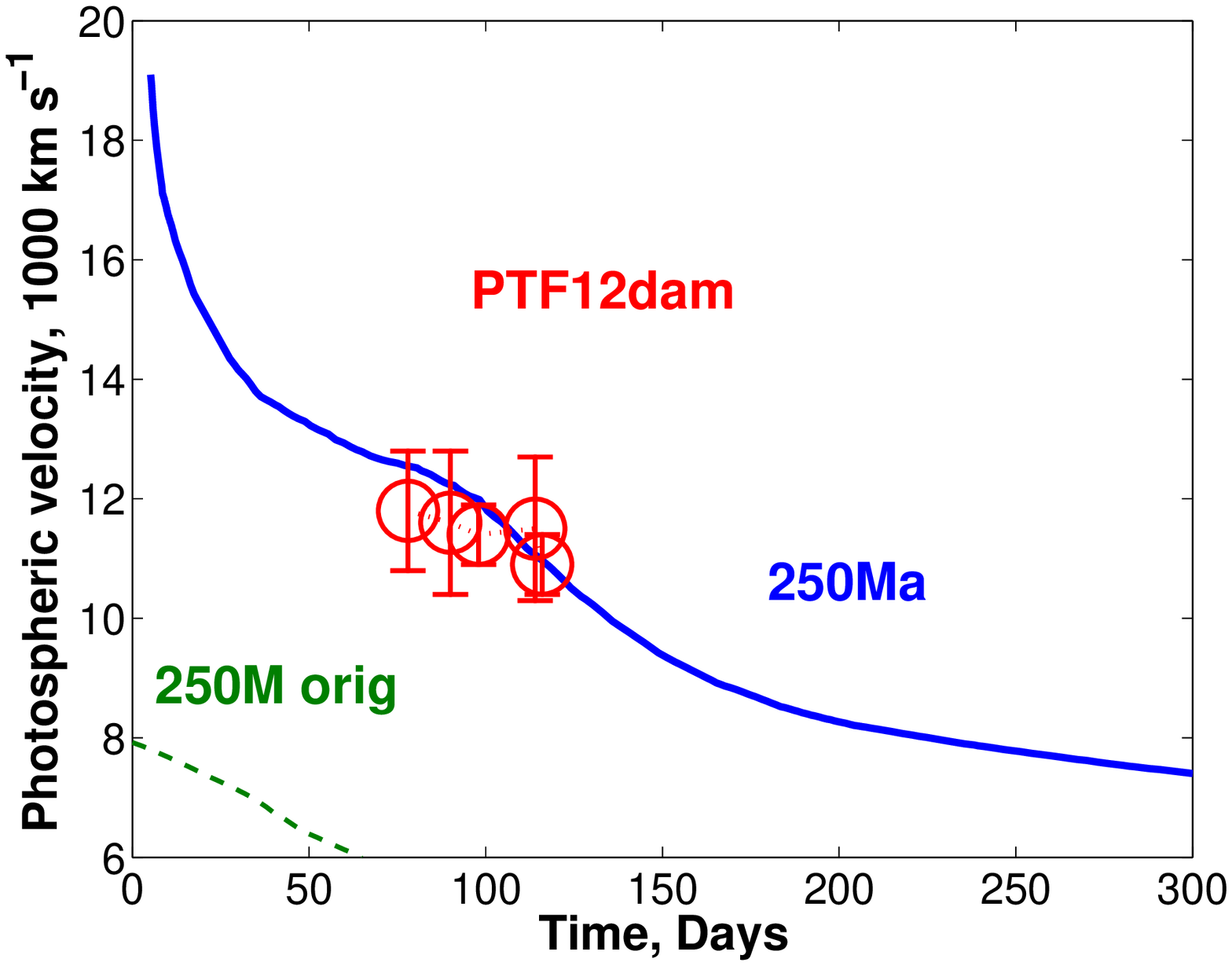}~
\includegraphics[width=0.5\textwidth]{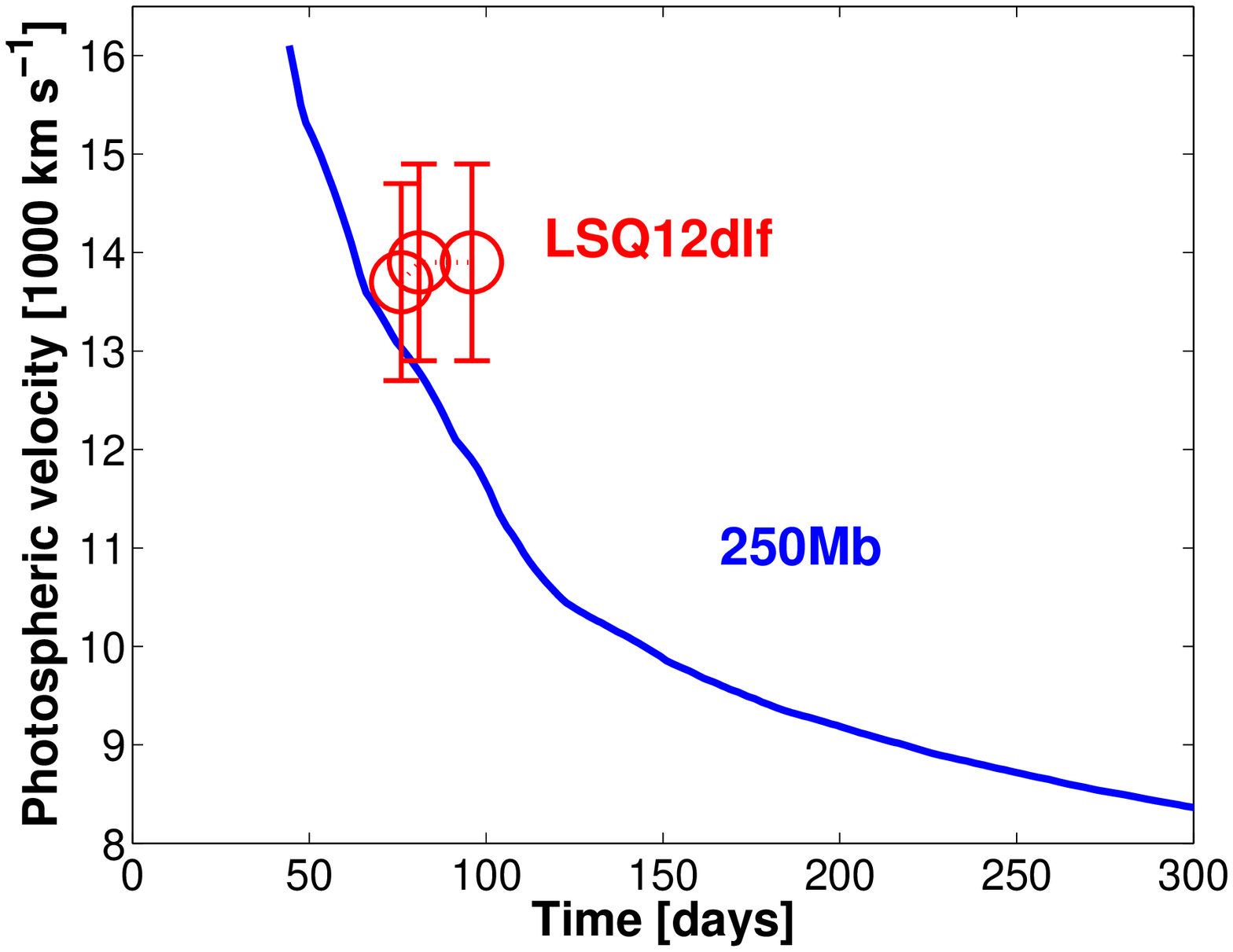}
\caption[Photospheric velocity of the original model 250M, mixed models 250Ma and 250Mb, and observed SLSNe
PTF12dam and LSQ12dlf]
{{\emph{Left}}: Photospheric velocity of the original model~250M (dashed
line), mixed model 250Ma and observed SLSN PTF12dam;
{\emph{Right}}: Photospheric velocity of the mixed model~250Mb and observed SLSN PTF12dam
\citep{2014MNRAS.444.2096N}.  See explanation in the text.}
\label{figure:Uph}
\end{figure*}

Through our simulations we show that a PISN model with abundance inversion might explain two particular SLSNe:
PTF12dam and LSQ12dlf.  Our qualitative fits are shown in Figures~\ref{figure:Lbol1}, \ref{figure:Lbol2},
\ref{figure:Tcol}, and \ref{figure:Uph}.

Model~250Ma traces the general behaviour of PTF12dam's light curve, with a
relatively sharp 50~day rise and post-maximum decline.  
Model~250Ma has colour temperature about 13000~K at the time of luminosity maximum.  The 250Ma temperature curve lies very close to 
the observed points.  Photospheric velocity at luminosity maximum is 12\,000~km\,s$^{\,-1}$ (Figure~\ref{figure:Uph}) and suitable
in matching the high photospheric velocity of PTF12dam.  
In principle, if our 250~\Msun{} PISN model possesses sufficient amount of nickel in the outer layers, then it might reproduce observational
appearance of SLSN~PTF12dam.

An alternative model is considered by \citet{2015AstL...41...95B}.  It is based on 
supernova shock interaction with circumstellar matter.  
The modelled light curves look suitable for PTF12dam.  However, the model requires
unrealistic input parameters, 5~\Msun{} helium ejecta colliding with the 100~\Msun{}
carbon--oxygen shell.  More realistic ejecta--circumstellar interaction model was suggested by
\citet{2014MNRAS.444.2096N} with
physically reasonable parameters (26~\Msun{} ejecta and 13~\Msun{} shell). 

Another semi-analytic magnetar-driven models for PTF12dam are available in the literature
\citep{2013Natur.502..346N,2014MNRAS.444.2096N,2015MNRAS.452.1567C}.  
The authors however think that the magnetar-powered supernova models 
encounter certain difficulties in explaining the high luminosity events
\citep{2015Badjin}.  In more detailed consideration,
magnetar rotational energy is not capable to be directly converted into thermal radiation.  
Thermal photon field causes high pair-creation opacity for gamma-photons in magnetar vicinity,
and hence prevent them to enter the expanding shell.  The spin-down energy
is converted into relativistic plasma pressure, and in turn into the kinetic energy of the inner shell.  
As a consequence, the resulting light curve does not reach luminosities required
by SLSNe.

We attempt to fit the narrower light curve of LSQ12dlf with the model~250Mb, in which we also shrink the hydrogen-helium atmosphere.  In general,
our model~250Mb fits the luminosity behaviour.  We suppose that, if the ejecta
retain even lower hydrogen-helium mass, then the
fit would be better.  However, through the present simulations we qualitatively demonstrate the opportunity for massive PISNe 
to explain fast SLSNe.  Modelled colour temperature and photospheric velocity of 250Mb
match those of LSQ12dlf.  

Below we discuss a few weak points which might arise for our ``macro-mixed'' PISN models.

\subsection{Hydrogen, helium and iron in SLSN spectra}

Many SLSNe are classified as supernovae~Type~Ic, because of the absence of hydrogen and helium in
their early spectra.


The absence or presence of helium in the spectra can be a clue point for macroscopic mixing.  
Similar ideas were discussed for SN~Ib and SN~Ic
explosions \citep[see e.g.,][]{1997ApJ...483..675C,2002ApJ...566.1005B,2003IAUS..212..346B,2012MNRAS.424.2139D}.  
As proposed, both SNe~Ib and SNe~Ic may have similar helium mass.  The difference between these two SN~types
arises from different amount of radioactive material mixed into the outer helium layers of the ejecta.  

Helium being mixed with radioactive material should be excited and emerge in the spectra.  
The minimum mass fraction of radioactive material is 0.01 as stated by \citet{2012MNRAS.424.2139D}.  
It is very difficult to hide hydrogen and helium if these species are microscopically mixed with
nickel, especially, if hydrogen-helium mass is very high, as PISN progenitors have \citep{2012MNRAS.422...70H}.  
If nickel is distributed in the form of macroscopic blobs without direct
mixing with hydrogen and helium, then it is likely that there is no
significant excitation by positrons from $\beta^{\,+}$--decay at
least at the earlier phase (S.~Hachinger, private communication).  Later on,
the blobs decay, and hydrogen/helium excitation might happen.

Numerous studies focusing both on observations and theoretical
simulations suggest that there are asymmetries in SN~explosions 
\citep[see e.g.,][and references
therein]{1997ApJ...483..675C,1999ApJ...521..179H,2006ApJ...645.1331M,2012MNRAS.424.2139D}.  
Asymmetric chemical distribution or clumping helps to avoid direct mixing of radioactive material with 
hydrogen and helium, which prevents non-thermal excitation and ionisation, and in turn avoids 
signatures of hydrogen/helium in the spectra.

In our mixed models, nickel is located in the outer layer.  During the
maximum phase this layer has temperature around 10000~K and 15000~K (models 250Ma and 250Mb
respectively).  At such a high temperature iron (created from nickel and cobalt) is ionised
(Fe\,III, Fe\,IV) and contributes only to far-UV, but not to visual light.

Overall, it remains controversial whether such a huge amount of
hydrogen and helium (20--50~\Msun{}) might be hidden or appear in the PISN spectra.

\subsection{The nature of mixing}

As we mention above, the displacement of chemical elements, if it happens, takes place simultaneously along 
with the pair-instability explosion.  
Macroscopic movement of stellar matter should proceed on the timescale of the explosion, i.e. dynamical time 
(fraction of second).  Convection operates on the longer characteristic time, governed by thermal 
Kelvin-Helmholtz scale, which is probably accelerated by the explosion in the central regions.  
The nickel-``core'' covers up to 60\,\% in radius of the entire exploding oxygen core, so that initialisation of 
chemical displacement could happen.  However,
we conclude that such macroscopic movement can more readily emerge from global anisotropy.  

Ignoring the great difference in the mass, the thermonuclear pair-instability explosion resembles
thermonuclear explosion of a white dwarf, which, as widely believed, results in SN~Ia.  
Both observations of SN~Ia \citep[see e.g.,][]{2014ApJ...784L..12G,2014ApJ...784...85P} and theoretical simulations
of white dwarf explosions \citep{2003Sci...299...77G} discover that during 
explosion of a white dwarf fast moving blobs from the 
innermost region penetrate the entire ejecta and emerge on the surface front.  Similarly, macroscopic replacement of stellar
matter might occur in the pair-instability explosion \textbf{inside} the exploding oxygen core.  However, as we show in
Section~\ref{subsect:halfres}, there are no suitable changes happening for the light curve evolution under this condition. 
If suddenly nickel is swept into the helium shell, and, for instance, the progenitor lost all
hydrogen, then the situation might result in a very different light curve. 

So far, all existing PISN evolutionary models are one-dimensional
\citep[][and others]{2002ApJ...567..532H,2002ApJ...565..385U,2013MNRAS.433.1114Y,2014A&A...566A.146K,2015ApJ...799...18C,2015ApJ...805...44S}.  
Even though some of these models are mapped into the
multi-dimensional code for following up the explosion, this could not produce any strong anisotropy
and macroscopic movement of ejecta matter
\citep[e.g.,][]{2014arXiv1407.7550C,2014ApJ...792...44C}.
Assuming strong perturbations in the entropy (or velocity) field during the earliest explosion phase might
result in high entropy plumes which in turn drives radioactive material from the innermost 
region into the upper layers.  Future numerical simulations will shed light on this question.

To summarise, until further studies determine if such strong mixing occurs,
the relevance of PISN models to SLSNe, especially rapidly rising examples,
remains unclear.

The recent results discover though that 
some very massive rotating stars at relatively high metallicity (0.001--0.002) lose all hydrogen and part of helium layers
\citep{2013MNRAS.433.1114Y,2014ApJ...797....9W,2015ApJ...799...18C}.  During the
pair-instability explosion some of the most massive and the most energetic models 
produce up to 40~\Msun{} of nickel, which is very
naturally, without any artificial modification, distributed up to the edge of the oxygen core.  The closeness of
radioactive material to the surface allows nickel to power the light
curve very soon after the shock-breakout event.  This makes PISNe more
suitable for SLSNe like PTF12dam.  The new results will be
described in a forthcoming paper.

\section{Conclusions}
\label{sect:conclusions}

In the present study we computed post-explosion photospheric evolution of pair instability supernovae.  
Our input models are based on the evolutionary model 250M \citep{2014A&A...566A.146K}.  
The original broad light curve resulting from the explosion of a very massive PISN progenitor makes 
PISNe difficult to interpret fast evolving SLSNe with relatively short rise and rapid decline in light curves.  
Nevertheless, \citet{2014A&A...565A..70K} show that at least some of
SLSNe (2007bi and PTF10nmn) might be easily explained by our 250~\Msun{}~PISN model.  

Strong mixing with extreme redistribution of metals along the ejecta
significantly modifies the observational appearance of PISN explosion.  
We made numerous attempts with different kinds of redistribution of species in
the original chemical structure of the 
250~\Msun{}~PISN model.  In particular, two toy models, 250Ma and 250Mb, appear the most suitable for SLSN
photospheric evolution.  In these models almost all radioactive nickel was relocated into the hydrogen-helium envelope.  
The resulting light curves have faster evolution than the original unmixed 250~\Msun{}~PISN model.  
The radioactive material in the upper layers provides the earlier maximum, shorter and steeper rise, and
higher colour temperatures and higher photospheric velocities around peak epoch.  
{{Other attempts involve mixing of nickel inside the oxygen core and 
totally mixed model.  These attempts do not result in a desirable
modification of the photospheric supernova evolution.}}

We compare our results to two SLSNe PTF12dam and LSQ12dlf.  The models~250Ma and 250Mb partly match their 
observed properties.
Hence, macro-scale clumping in the PISN ejecta or strong anisotropy might help PISN to be a good candidate for SLSNe.  
Mixing of 20~\Msun{} of nickel (even in blobs) with 25--50~\Msun{} of hydrogen and helium would
likely provide spectral signatures by non-thermal excitation mechanism. 
However, it is not clear how strong might be the effect of excitation.  On top of that,
realisation of such an extreme chemical redistribution is problematic.  
Therefore, we emphasise that 
our hydrogen-helium rich massive PISN model are not relevant for explaining 
main features of hydrogen-helium poor rapidly rising SLSNe.  

Looking differently, our results might mean that PISN ejecta need to be
``mixed'' to explain SLSN~observations.  In addition, the observed nebular spectra of
SLSN~2007bi and some other SLSNe also may require some mixing in the SN~ejecta
\citep{2015AJerkstrand}.  
In particular, oxygen lines are narrow and iron lines are broad
\citep[see e.g.,][]{2009Natur.462..624G,2014MNRAS.444.2096N}, which, in turn, means that
at least some amount of oxygen is located close to the centre, and iron is
located far from the central region.

To conclude, we stress that PISN originating from the explosion of very massive star is not the best candidate
for majority of SLSNe.  Nevertheless, rotating PISN models might be more suitable at least for some
slowly-rising SLSNe.  
Future stellar evolutionary calculations, multi-dimensional simulations of
the pair-instability explosion and supernova simulations will shed light on this question.  


\section*{Acknowledgements}

AK acknowledges support from EU-FP7-ERC-2012-St Grant~306901.  
AK highly appreciates all comments and ideas emerged during extensive 
discussions with Takashi Moriya, Norbert Langer, Debashis Sanyal, Anders Jerkstrand, Matt Nicholl,
Steven Smartt, Mikhail Basko, Ken'ichi Nomoto, Avishay Gal-Yam, and
Stephan Hachinger.  
SB is supported by a grant from the Russian Science Foundation (project
number~14-12-00203).
\addcontentsline{toc}{section}{Acknowledgements}

\bibliographystyle{mnras}

\bsp	
\label{lastpage}
\end{document}